# Agentic AI for IPoDWDM Network Lifecycle Automation: An MCP-Enabled Architecture

Chunmin Xia, Jakub Harbaczewski, Nikhil Dsilva, Julie Raulin, Dominic Schneider, Achim Autenrieth

Adtran Networks SE, chunmin.xia@adtran.com

**Abstract** *We present a distributed, vendor-agnostic multi-MCP architecture for SDN-based automation and autonomous control of multi-vendor, multi-layer IPoDWDM networks. The framework enables E2E service lifecycle automation, closed-loop cross-layer control using GNPy model and optical telemetry, and is experimentally validated on a IPoDWDM testbed.* 

## Introduction

The rapid development and growing applications of artificial intelligence (AI), together with the explosive growth of different data hungry applications, is accelerating requirements for highly dynamic, deep automation and increasing autonomy imposed on IP-over-DWDM (IPoDWDM) networks. As an universal, standardized interface between autonomous AI agents and the real-world systems, model context protocol (MCP) [1,2] enables scalable and composable development and integration of different kinds of tools in different specific domains. Agentic AI application concepts focusing on some specific areas are seen in [3-9]. Based on both simulator/digital twin and real-world test beds, we propose and experimentally demonstrate an MCP-server-based agentic-AI architecture that enables fully automated, vendor-agnostic, multi-layer lifecycle management of IPoDWDM networks, spanning physical-layer setup through IP-layer provisioning and performance telemetry monitoring.

## Network setup and assumptions

As shown schematically in Fig. 1, we consider a representative ring network with at least three network nodes, each comprising ROADMs, variable gain booster and pre-amplifiers as well as couplers/filters, interconnecting coherent pluggables from routers of different vendors including Juniper, Cisco and Arista as well as transponders from Adtran. It is necessary to point out our MCP servers can be easily extended to support any other routers/transponders with given necessary NETCONF configuration files. The Mosaic Network Controller (MNC), an SDN controller from Adtran, is used as the SDN control platform and is compatible with the available hardware testbed.

## AI agent and MCP tools development

We have developed our own AI agent based on a Large Language Model (LLM) GPT5.2. The client prompts are saved in a memory datastore using Redis and attached to the Agent together with any new requests to keep the memory of the context within one session. By evaluating streamable HTTP-based MCP servers and our own agent against commercial agents (e.g., Copilot, Claude), we identified limitations and iteratively improved our MCP servers and agent.

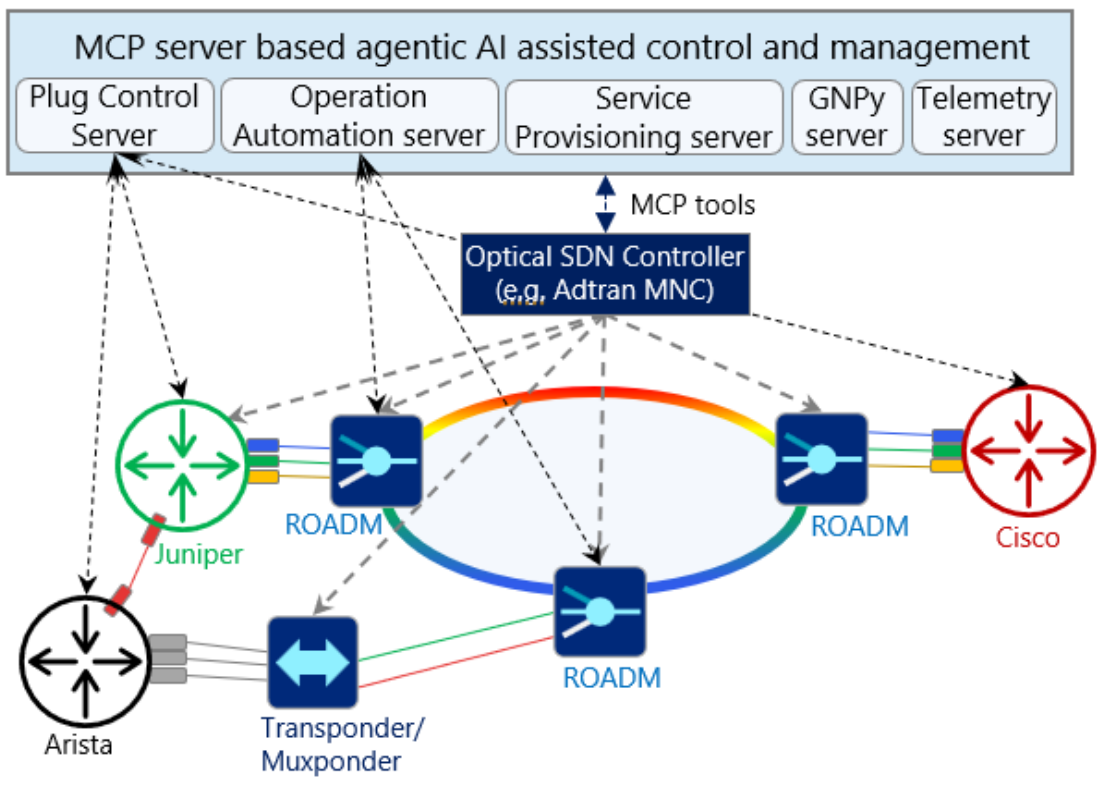

**Fig. 1: Agentic AI assisted IPoDWDM network based on multiple MCP servers**

In considering the importance of security and trustworthiness along with the AI applications, we keep always the harness engineering discipline in mind during developing all the MCP servers. First, all authentication credentials are managed through environment-based configuration parameters for general use cases, while policy-driven Eclipse Dataspace Components Connector (EDCC) [10,11] mechanisms are used to meet higher-level security requirements. Secondly, in order to improve the transparency and ensure a positive user experience, each MCP tool is divided into as many atomic steps as possible, each of which is checked or validated carefully in order to return a clear meaningful information for any conditions even when an unexpected failure occurs.

The MCP servers are developed and tested first of all based on digital twin like Adtran network simulators and then validated in real networks.

## Multi-MCP server architecture and workflow

We aim to realize a deep smart automation of IPoDWDM network with human language to AI Agent through creating different MCP servers working jointly in multiple network operation layers from physical equipment setup, fiber connections, device configuration to service provisioning and performance estimation as well as telemetry. As shown in Fig. 2, we have implemented and tested the following MCP servers:

**1. Operation automation server**: by a human prompt with a given SDN controller IP, any sub-network topology can be loaded automatically into a webpage. Nodes, shelves, cards and ports are all visualized for any further operations including fiber or nodes connections, device configurations and service creation, either through the GUI or the prompt or the combination of both. As indicated by the top part in Fig. 2, the user can prepare network operational data (e.g., fiber connections) via the GUI, visually review it, save it into batch files, and submit them to the Agent for processing with one prompt. All the creation- or deletion-requests can be verified by the corresponding show tool.

**2. Coherent pluggables control server**: Coherent pluggables in routers or transponders are controlled directly by *plug control MCP server*, which realizes the following three fundamental functions: configuration, configuration status check or verification and performance metric monitoring after provisioning end-to-end (E2E) service. The detailed workflow is shown Fig. 3.

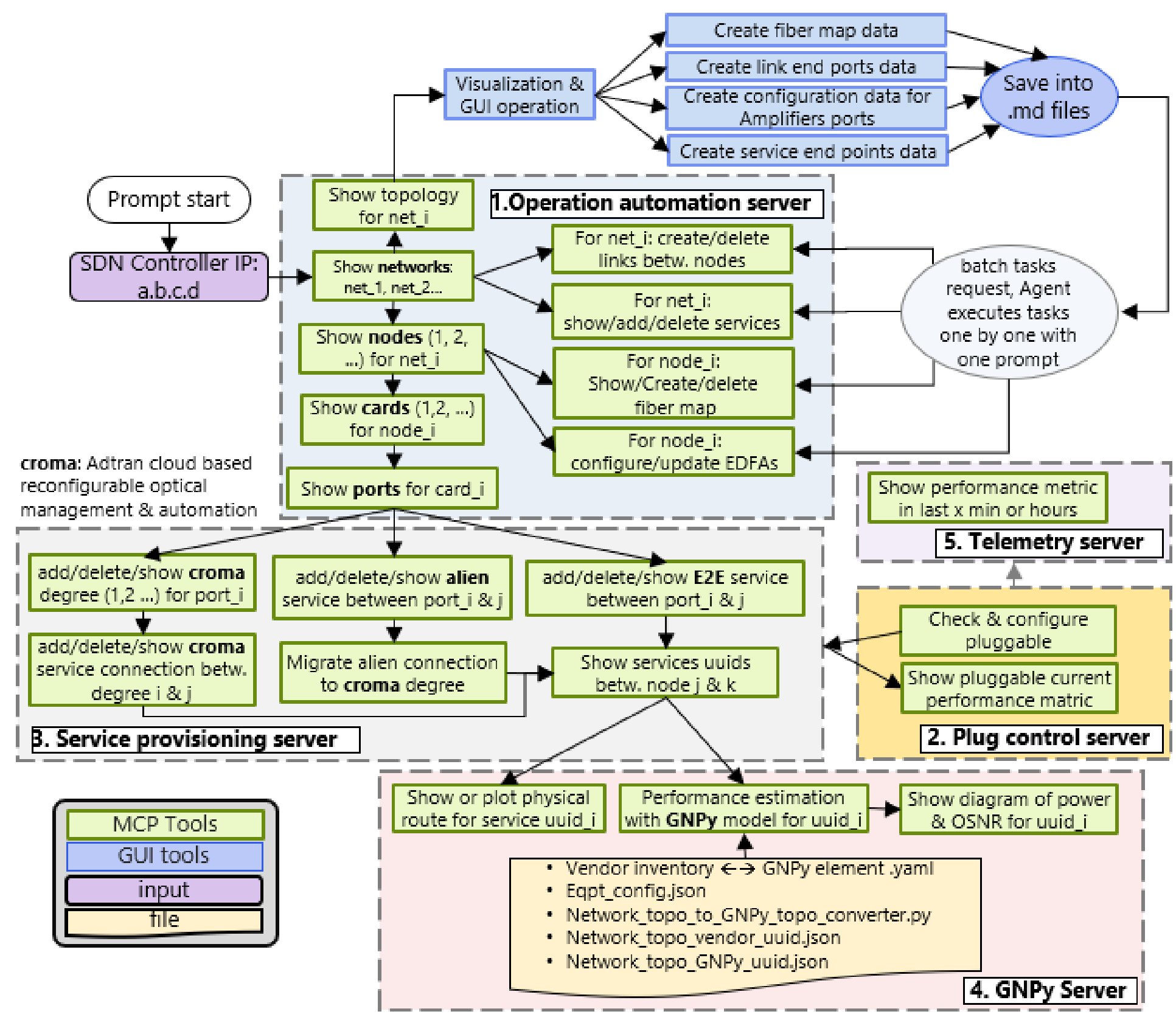

**Fig. 2:** Workflow chart of MCP servers enabled IPoDWDM network

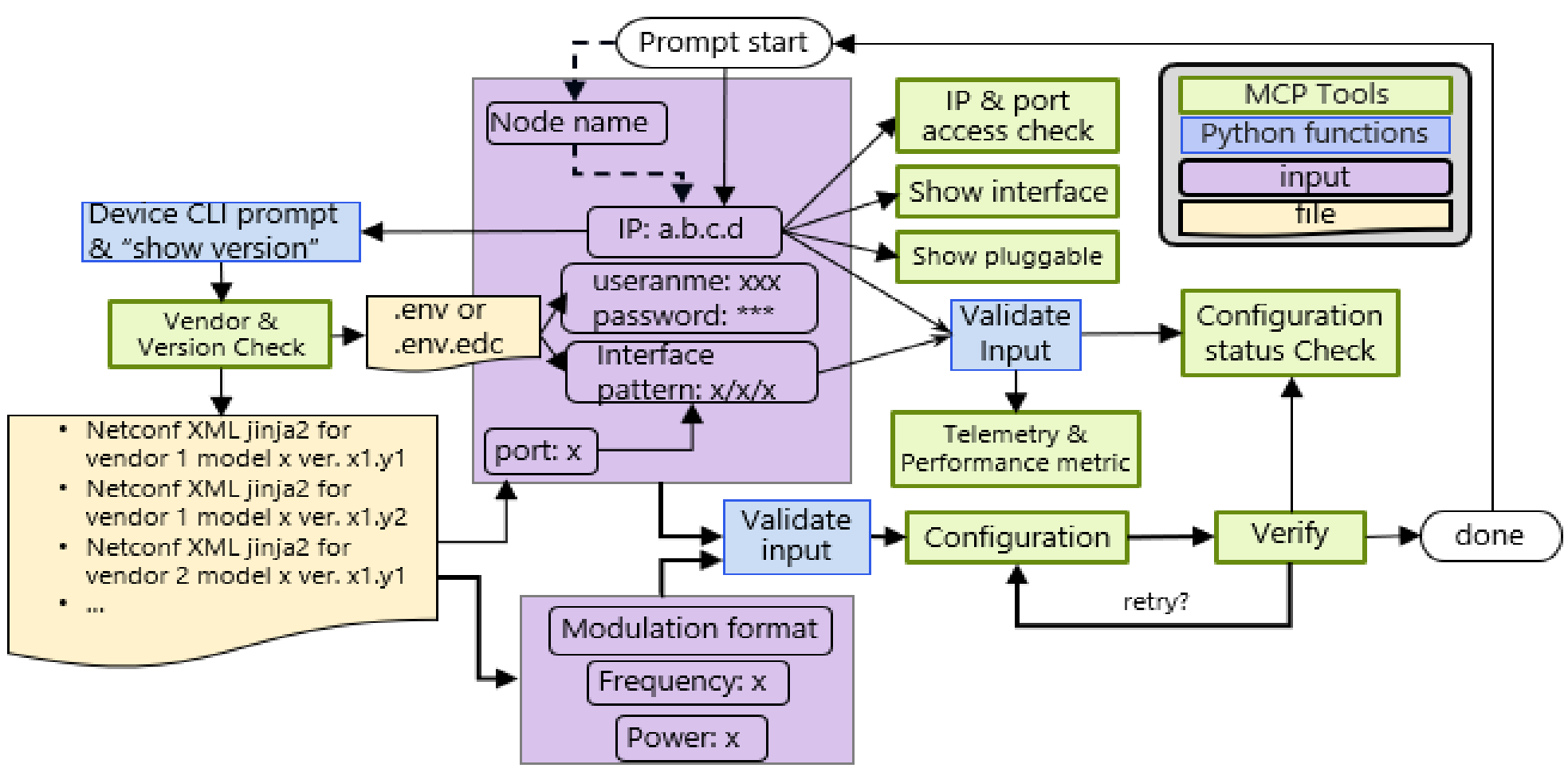

**Fig. 3:** Workflow chart of coherent plug control MCP server

An important observation is that the current state of the art of routers and pluggable optics from different vendors remains heterogeneous with respect to their support for NETCONF, OpenConfig, and gNMI protocols. Even with a same router vendor, the concrete NETCONF /openconfig XML RPC templates may be variable for different router models or different operating system (OS) software versions. Consequently, vendor information, hardware models, and OS versions were obtained via NETCONF or CLI and mapped to or inserted into vendor-specific Jinja2-based NETCONF XML templates (Fig. 3), enabling vendor-agnostic management by hiding vendor-specific differences from the agent.

**3. Service provisioning server**: Three different services provisioning MCP tools have been implemented: (1) based on Adtran cloud-based reconfigurable optical management & automation (croma) platform by creating the service connections step by step from one ROADM degree to another. This is considered as a generic approach and does not need SDN controller support. (2) Create an alien service first between the ports from the OLS side that are connected to the router ports. (3) Single-step E2E service provisioning is currently supported on Juniper routers, with ongoing implementation to extend compatibility to Cisco and Arista devices.

**4. GNPy server**: after service provisioning, GNPy server is used to estimate and predict the E2E performance. As shown in Fig. 2, in order to use the open source GNPy model [12], all the network elements together with their configuration and spec parameters in the service path from the real network topology and equipment configurations are firstly discovered and then converted into GNPy model compatible network .json format. Then the power and OSNR evolution along the path can be calculated.

**5. Telemetry server**: All the performance metrics of coherent transceivers can be retrieved based on queried data set and history time slot.

## Validation and demonstration

Based on a real testbed shown in Fig. 4(a) and (b) by user prompts, two 400G DP-16QAM coherent optical pluggables (operated in 400ZR DWDM amplified mode) from the two Juniper routers (ptx10001) are amplified (not shown in (b)) and then connected to a 16x4 filter, followed by a 12-degree ROAMD (RD-12RS) equipped with booster and pre-amplifier (AM_S23L). One SSMF spool of 100km between node F8_41and F8_45. Rather than listing all network operation results, Fig. 4(c) highlights the GNPy-based performance estimation for the example service path. For reference, the OSNR including nonlinear interference (NLI) under full C-band loading is shown alongside the measured OSNR for single-channel transmissions at 191.5 , 193.1, and 196 THz. The performance metrics of the two coherent transceivers are reported in Fig. 4(d).

## Conclusions

We proposed an Agentic AI orchestrated collaboratively multi-MCP server architecture enabling full and efficient SDN-based automation. Experimental validation on a real IPoDWDM testbed demonstrates the feasibility to streamline E2E service lifecycle operation and management across physical, optical and IP layers.

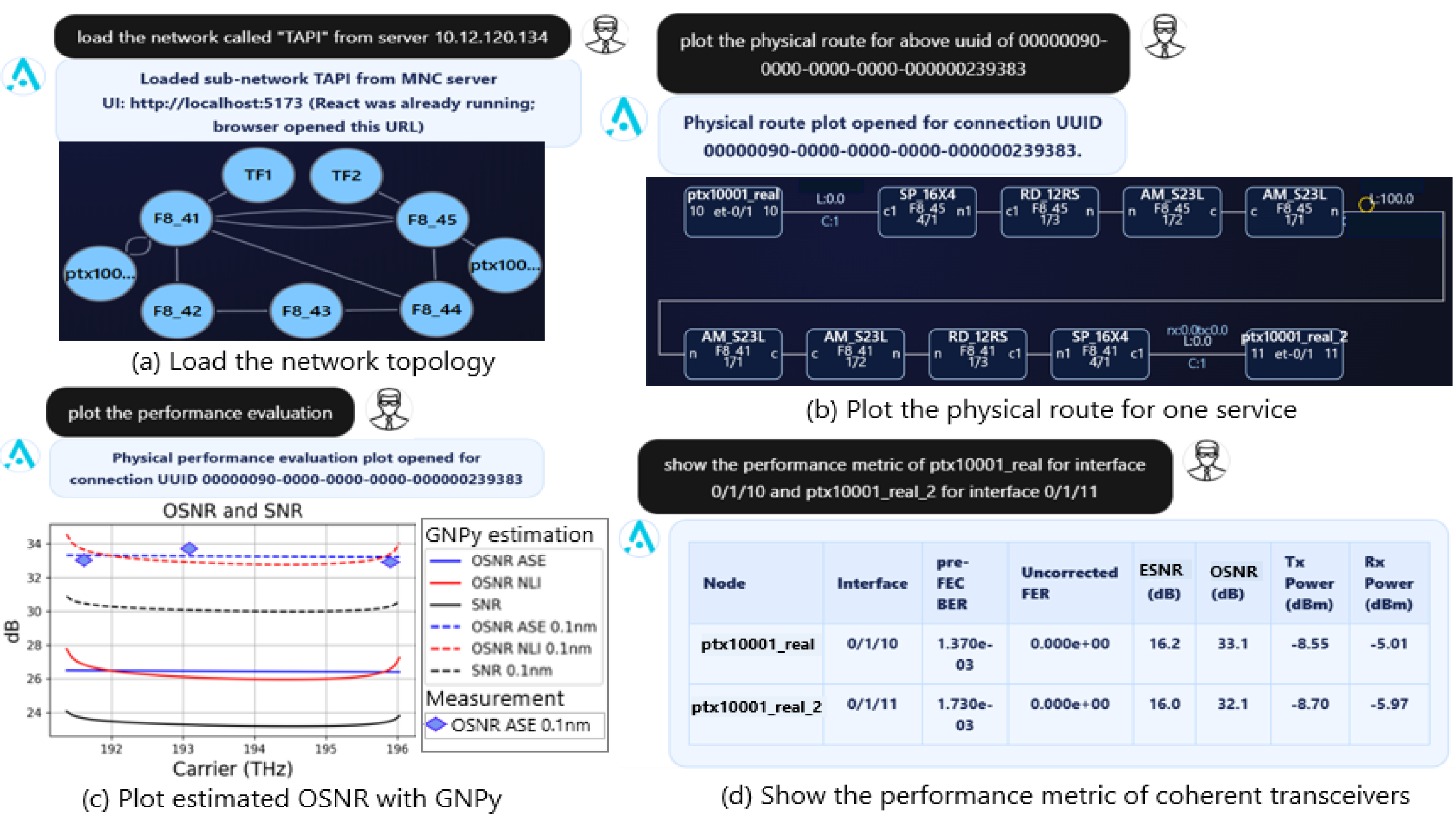

| Node | Interface | pre-FEC BER | Uncorrected FER | ESNR (dB) | OSNR (dB) | Tx Power (dBm) | Rx Power (dBm) |
|---|---|---|---|---|---|---|---|
| ptx10001_real | 0/1/10 | 1.370e-03 | 0.000e+00 | 16.2 | 33.1 | -8.55 | -5.01 |
| ptx10001_real_2 | 0/1/11 | 1.730e-03 | 0.000e+00 | 16.0 | 32.1 | -8.70 | -5.97 |

(a) Load the network topology

(b) Plot the physical route for one service

(c) Plot estimated OSNR with GNPy

(d) Show the performance metric of coherent transceivers

Fig. 4: Sample MCP outputs from a real IPoDWDM network testbed

## Acknowledgment

This work has received funding from the German Federal Ministry of Research, Technology, and Space (BMFTR) project SUSTAINET-Advance, Grant 16KIS2271K, in the framework of the CELTIC-NEXT project id C2024/3-3.